\documentstyle[prd,aps,eqsecnum]{revtex}

\begin{document}

\title{Transition Temperature for Weakly Interacting Homogeneous 
Bose Gases}

\author{Frederico F. de Souza Cruz\thanks{Email address: fred@fsc.ufsc.br},
Marcus B. Pinto\thanks{Email address: fsc1mep@fsc.ufsc.br }}

\address{
{\it Departamento de F\'{\i}sica,}
{\it  Universidade Federal de Santa Catarina,}\\
{\it 88040-900 Florian\'{o}polis, SC, Brazil}}

\author{Rudnei O. Ramos\thanks{Email address: rudnei@peterpan.dartmouth.edu.
$^{(1)}$Permanent address.}} 

\address{
{\it Department of Physics and Astronomy, Dartmouth College,}
{\it  Hanover, New Hampshire 03755-3528}\\
{\it and $^{(1)}$Departamento de F\'{\i}sica Te\'orica,}
{\it Instituto de F\'{\i}sica,}\\
{\it Universidade do Estado do Rio de Janeiro, }
{\it 20550-013 Rio de Janeiro, RJ, Brazil}}

\maketitle

\begin{abstract}

We apply the non-perturbative optimized linear $\delta$ expansion method
to the $O(N)$ scalar field model in three-dimensions to determine the
transition temperature of a dilute homogeneous Bose gas. Our results
show that the shift of the transition temperature, $\Delta T_c/T_c$, of
the interacting model, compared with the ideal gas transition
temperature, really behaves as $\gamma a n^{1/3}$ where $a$ is the
s-wave scattering length and $n$ is the number density. For $N= 2$ our
calculations yield the value $\gamma= 3.059$. 

\vspace{0.34cm}
\noindent
PACS number(s):  03.75.Fi, 05.30.Jp, 11.10.Wx

\end{abstract}

\section{Introduction}

The experimental realization of the Bose-Einstein condensation in dilute
atomic gases has greatly stimulated an enormous number of theoretical
studies in this field (for recent reviews on the theory and experiments, see
for instance, Refs. \cite{dalfovo,other}). Most of this interest comes from
the fact that in these experiments a great deal of control can be achieved
in almost every parameter of the system. Thus, experiments in dilute Bose
gases provide a perfect ground to test numerous models and ideas, as for
example those commonly used in quantum field theory, applied to
non-relativistic systems. In particular, a theoretical study which has
attracted some attention very recently is the determination of the behavior
of the transition temperature in the presence of a repulsive interaction.
This non trivial problem has been treated by different methods with
different results. Taking a dilute Bose gas, with a repulsive interaction
characterized by the scattering length parameter $a$, the dependence on $a$
for the difference between the critical temperature shift with and without
interaction ($\Delta T_c/T_{c}=(T_{c}-T_{c}^{0})/T_{c}$) is a highly
controversial point, be with respect to the functional dependence on $a$ or
even regarding the sign. An early Hartree-Fock calculation \cite{fetter} with
a non-delta interaction gave a negative sign for $\Delta T_c$. This very same
sign was also obtained by Toyoda \cite{toyoda} using one loop
renormalization group and obtaining a functional dependence on $a$ as
$\Delta T_c/T_{c}=\gamma \left( a^{3}n\right) ^{1/6}$,
where $n$ is the density. More recently, Huang \cite{huang} obtained the
same dependence for $\Delta T_c/T_{c}$, but with a positive constant
$\gamma$. 
Gr\"uter, Ceperley and Lalo\"e \cite{gruter},
Holzmann and Krauth \cite{holzmann-krauth} and 
Holzmann, Gr\"uter and Lalo\"e \cite{markus} investigated the dependence of 
$\Delta T_c$ numerically using Monte Carlo methods. They obtained, in the low
density limit, a dependence of the type $\Delta T_c/T_{c}=\gamma \left(
a^{3}n\right)^{1/3}$ but with different values for $\gamma$. 
More recently the Monte Carlo technique has been again applied to 
this problem by Prokof'ev and Svistunov \cite {russos} and Arnold and 
Moore \cite {arnold1}. These authors, who obtain $\gamma=1.29\pm 0.05$ and 
$\gamma=1.32\pm 0.02$, respectively, claim that their results are
more accurate than those obtained in 
Refs. \cite{gruter,holzmann-krauth,markus}.

One of the reasons
for the multitude of results and methods stem from the fact that at
the transition temperature  ordinary perturbation theory fails 
(due to infrared divergences) and we must resort to non-perturbative 
methods. 
Recently, there have been also used various non-perturbative methods 
to treat the problem of
the transition temperature from an analytical way. {}For example,
the authors in 
Ref. \cite{baymprl} perform a self-consistent calculation obtaining 
$\Delta T_c/T_c = \gamma a n^{1/3}$, with $\gamma \simeq 2.9$. 
The non-perturbative $1/N$ method  has been also used to determine the shift. 
Its leading order contribution has been evaluated by Baym, Blaizot 
and Zinn-Justin \cite{baymN} 
who obtain $\Delta T_c/T_c = \gamma a n^{1/3}$, with 
$\gamma \simeq 2.33$ for $N=2$. Considering the next to leading order
term, Arnold and Tom\'asik \cite{arnold} determine a correction to this 
large-$N$ expansion, obtaining a value for $\Delta T_{c}/T_{c}$ which
is $\sim 26 \%$ smaller. 

The  results obtained by Bijlsma and 
Stoof \cite{stoof}, who used renormalization group techniques,
and Baym  et al. \cite{baymprl} 
were compared  with the temperature transition  data in the Vycor-$^{4}He$ 
system by Reppy et al. \cite{reppy}. Those  experiments 
seem to give a somewhat larger value for the
constant $\gamma$, as $\gamma \sim 5$ \cite{reppy}. This value is close
to $\gamma \simeq 4.66$ which is the one obtained in Ref. \cite{stoof}. 
Nevertheless some authors argue that 
this system would not exactly correspond to a dilute Bose gas of hard 
spheres \cite{arnold}. 

In this paper we apply the non-perturbative linear $\delta$ expansion 
(or optimized
perturbation theory) \cite{linear} (for earlier references see, {\it
e.g.}, Refs. \cite{seznec,seznec1,seznec2}) to an effective model for dilute 
homogeneous Bose gases. 
This approximation has been shown
to be a powerful non-perturbative method, and sufficiently simple to use
in very different applications, including the study of non-perturbative
high temperature effects, as shown very recently in the context of
finite temperature quantum field theory \cite{MR1} as well as finite
chemical potential \cite {hugh}. The method also introduces an arbitrary mass
parameter which prevents infrared divergence problems. Non-perturbative
results are generated when one optimizes the theory with respect to this
mass parameter. 

The paper is organized as follows. We present the method in Sec. II 
illustrating with an application to the 
pure anharmonic oscillator which has many similarities with the model used 
here 
to describe dilute Bose gases. The interpolated version of an effective model 
for weakly interacting homogeneous Bose gases is obtained in Sec. III. 
Sec. IV is devoted to the perturbative evaluation of density related 
quantities. In Sec. V we present our non-perturbative results for the 
critical temperature shift comparing with results available in the 
literature. The conclusions are presented in Sec. VI. 

\section {The linear $\delta$ expansion}

The optimized linear $\delta$ expansion (LDE) is an alternative 
non-perturbative approximation which has been successfully used in a plethora 
of different problems in 
particle theory 
\cite{linear,okotem,landau,njlft}, quantum mechanics 
\cite {seznec1,ian,guida}, nuclear matter \cite {gas}, lattice 
field theory \cite{evans} as well as for determining the equation of state 
for the Ising model \cite {zj}.
One advantage of this method is that 
the selection and evaluation (including renormalization) of Feynman diagrams 
are done exactly as in perturbation theory using a very simple modified 
propagator which depends on an arbitrary mass parameter. The results are 
optimized with respect to this parameter at the end of the calculation. The 
standard application of the LDE to a theory described by some Lagrangian 
density 
${\cal L}$ starts with an interpolation defined by

\begin{equation}
{\cal L}_{\delta} = (1-\delta){\cal L}_0(\eta) + \delta {\cal L} = 
{\cal L}_0(\eta) + \delta [{\cal L}-{\cal L}_0(\eta)],
\label{int}
\end{equation}

\noindent 
where ${\cal L}_0(\eta)$ is the Lagrangian density of a
solvable theory which can contain arbitrary mass parameters $\eta$.
The Lagrangian density ${\cal L}_{\delta}$ interpolates between the
solvable ${\cal L}_0(\eta)$ (when $\delta=0$) and the original ${\cal
L}$ (when $\delta=1$). To illustrate how the method works let us consider 
the anharmonic oscillator described by

\begin{equation} 
{\cal L} = 
\frac{1}{2} (\partial_{0}\phi)^2 - \frac{1}{2} m^2
\phi^2 - \frac {\lambda}{4} \phi^4 \;.
\label{model} 
\end{equation} 
{}Following the interpolation prescription given by Eq. (\ref {int}) 
one may choose
 
\begin{equation} 
{\cal L}_0(\eta) =  \frac{1}{2}
(\partial_{0}\phi)^2 - \frac{1}{2} m^2 \phi^2 - \frac{1}{2}
\eta^2 \phi^2  \;,
\end{equation}
obtaining
\begin{equation}
{\cal L}_{\delta} = \frac{1}{2} 
(\partial_{0}\phi)^2 - \frac {1}{2} \Omega^2 \phi^2 
-\delta \frac {\lambda}{4} \phi^4 + \frac {\delta}{2} 
\eta^2 \phi^2 \;,
\label{eqdel}
\end{equation}

\noindent
where $\Omega^2=m^2+\eta^2$.
The general way the method works becomes clear by looking at the Feynman
rules generated by ${\cal L}_{\delta}$. {}First, the original $\phi^4$
vertex has its original Feynman rule $-i 6\lambda$ modified to $-i6\delta
\lambda$. This minor modification is just a reminder that one is really
expanding in orders of the artificial parameter $\delta$. Most
importantly, let us look at the modifications implied by the addition of
the arbitrary quadratic part. The original bare propagator, 

\begin{equation}
S(k)= i(k^2-m^2 +i\epsilon)^{-1}\;, 
\end{equation}

\noindent
becomes
\begin{equation}
S(k)= i(k^2-\Omega^2 +i\epsilon)^{-1}=
{i \over {k^2 - m^2 + i\epsilon  }}\left[ 1 - 
{\frac{i}{k^2-m^2 +i\epsilon} (-i\eta^2)}
\right ]^{-1}\,,
\label{prop}
\end{equation}

\noindent
indicating that the term proportional to $ \eta^2 \phi^2$ contained in
${\cal L}_0$ is entering the theory in a non-perturbative way. On the
other hand, the piece proportional to $\delta\eta^2 \phi^2$ is only
being treated perturbatively as a quadratic vertex (of weight $i \delta
\eta^2$). Since only an infinite order calculation would be able to
compensate for the infinite number of ($-i\eta^2$) insertions contained
in Eq.~(\ref {prop}), one always ends up with a $\eta$ dependence in any
quantity calculated to finite order in $\delta$. Then, at the end of the
calculation one sets the dummy parameter $\delta$ to unity  (the value at 
which the original theory
is retrieved) and fixes $\eta$ with the variational procedure known as
the Principle of Minimal Sensitivity (PMS) \cite {pms} which requires that 
a physical quantity $P$ calculated 
{\it perturbatively} in 
powers of $\delta$ be evaluated at the point where it is less sensitive to 
variations of the arbitrary $\eta$. That is, one optimizes the perturbative 
calculation by requiring
\begin{equation}
\frac{\partial P(\eta)}{\partial {\eta}} \Biggl |_{\bar {\eta}} =0\;.
\label{pms1}
\end{equation}
This procedure gives $\bar{\eta}$ as a function of the original parameters, 
including the couplings, and generates non-perturbative results as shown in 
the numerous applications cited above.
 
As a warm-up for our application to the Bose gas problem we follow Bellet, 
Garcia and Neveu \cite{phil} evaluating the ground state energy density 
${\cal E}$ and the vacuum expectation value $\langle \phi^2 \rangle$ for 
the anharmonic oscillator.  Other applications to this problem can be found 
in Refs. \cite{seznec1,ian,guida}.

By taking $m=0$ in Eq. (\ref{model}) one obtains 
the Lagrangian density for the pure anharmonic oscillator (PAO) which cannot 
be treated by ordinary perturbation theory. The exact result 
\begin{equation}
{\cal E}^{\rm exact} = \lambda ^{1/3}\;0.420804974478 \ldots
\end{equation}
has been obtained by Bender, Olaussen and Wang \cite{carl},
while Banerjee et al. \cite{bane} have obtained the exact result for 
$\langle \phi^2 \rangle$ 
\begin{equation}
 \langle \phi^2 \rangle^{\rm exact} = \lambda^{-1/3} \; 0.456119955748...
\end{equation}
In quantum field theory, the ground state energy density is represented by 
vacuum to vacuum diagrams. The relevant contributions to 
${\cal O}(\delta^2)$ are \cite {phil}
\begin{eqnarray}
{\cal E}^{(2)}(\eta)&=& -\frac{i}{2}\int_{-\infty}^{+\infty} 
\frac {dp}{2\pi} \ln \left [p^2-\Omega^2 \right ] -\delta \frac{i}{2}
\int_{-\infty}^{+\infty} \frac {dp}{2\pi}\frac{\eta^2}{p^2-\Omega^2}
\nonumber \\
&-&\delta\lambda \frac{3}{4}\left (\int_{-\infty}^{+\infty} 
\frac {dp}{2\pi}\frac {1}{p^2-\Omega^2}\right )^2 +\delta^2 
\frac{i}{4}\int_{-\infty}^{+\infty} \frac {dp}{2\pi}
\left [ \frac {\eta^2}{p^2-\Omega^2} \right ]^2 \nonumber \\
&+& \delta^2 \lambda\frac{3}{2}\int_{-\infty}^{+\infty} 
\frac {dp}{2\pi}\frac {1}{p^2-\Omega^2}\int_{-\infty}^{+\infty} 
\frac {dp}{2\pi}\frac {\eta^2}{(p^2-\Omega^2)^2} \nonumber \\
&-&\delta^2\lambda^2 \frac{9}{4}\left (\int_{-\infty}^{+\infty} 
\frac {dp}{2\pi}\frac {1}{p^2-\Omega^2}\right )^2 \int_{-\infty}^{+\infty} 
\frac {dp}{2\pi}\frac{1}{(p^2-\Omega^2)^2} \nonumber \\
&-&\delta^2 \lambda^2\frac{3}{4}\int_{-\infty}^{+\infty} \frac {dp}{2\pi}
\int_{-\infty}^{+\infty} \frac {dq}{2\pi}\int_{-\infty}^{+\infty} 
\frac {dl}{2\pi}\left [ \frac{1}{(p^2-\Omega^2)(q^2-\Omega^2)(l^2-\Omega^2)} 
\right . \nonumber \\
&\times&\left . \frac{1}{(p+q+l)^2-\Omega^2} \right ] + O(\delta^3)\;\;.
\end{eqnarray}

\noindent
Note that the second, fourth  and fifth 
contributions are due to the extra quadratic vertex, while all the others 
would 
also appear in an ordinary perturbative expansion to ${\cal O}(\lambda^2)$. 
Setting $m=0$ (PAO), evaluating the integrals and eliminating the divergent 
${\cal O}(\delta^0)$ term one obtains
\begin{equation}
{\cal E}^{(2)}(\eta)= {\cal E}^{(1)}(\eta)+\delta^2\frac{3\lambda}{16\eta^2}
-\delta^2 \frac{21\lambda^2}{128 \eta^5} \;,
\end{equation}
where
\begin{equation}
{\cal E}^{(1)}(\eta)= \delta \frac{1}{4}\eta + \delta\frac{3\lambda}{16\eta^2}
\;\;.
\end{equation}

Good numerical results appear already at first order where the application 
of the PMS  to ${\cal E}^{(1)}(\eta)$ yields ${\cal E}^{(1)}({\bar \eta})
=\lambda^{1/3}\; 0.4290$ at ${\bar {\eta}}=(\lambda \; 1.5)^{1/3}$.
To second order the authors in Ref. \cite {phil} obtain ${\cal E}^{(2)}
({\bar \eta})=\lambda^{1/3} \; 0.4210$ and then carry on improving this 
result to show 
convergence. The interested reader is referred to Ref. \cite {phil}  for 
details concerning the optimization procedure (selection of roots, etc). 
Other proofs of convergence are given in Refs. \cite {ian} and \cite {guida}.

Bellet, Garcia and Neveu also investigate the vacuum expectation value 
$\langle \phi^2 \rangle$ and we discuss their results here because this 
physical quantity is particularly important for our application to bosonic 
condensates. The perturbative expansion for $\langle \phi^2 \rangle$ can be 
obtained in different ways using standard quantum field theory methods. 
The authors in Ref. \cite{phil} prefer to do it from the perturbative 
expansion for ${\cal E}^{(\delta)}$ recalling that

\begin{equation}
\langle \phi^2 \rangle^{(\delta)}= 2 \frac{\partial}{\partial \Omega^2} 
{\cal E}^{(\delta)} \;.
\end{equation}
Going to second order in $\delta$ they optimize this quantity in two 
different ways. {}First, by applying the PMS condition directly to 
$\langle \phi^2 \rangle^{(2)}$ they obtain $\langle \phi^2 \rangle^{(2)}=
\lambda^{-1/3} \;0.455758$. Next, they use the optimum values obtained by 
extremizing ${\cal E}^{(2)}$, getting $\langle \phi^2 \rangle^{(2)}=
\lambda^{-1/3} \;0.454246$, showing that both approaches lead to results 
with same order of accuracy.

Still in the context of the anharmonic oscillator, Jones, Parkin and
Winder in Ref. \cite{aodin} have shown that the linear $\delta$ expansion 
applied to the calculation  of dynamical evolution of $\langle \phi^2 \rangle$,
where the PMS is applied directly to this quantity, tracks the exact solution
longer than any previous approximate methods used to study the same 
quantity, like Hartree-Fock or ordinary perturbation theory. This result
also reinforces the correctness of our procedure of optimizing
the density in this particular application to Bose condensates.

Before applying the $\delta$ expansion to the Bose-Einstein condensation 
problem let us clarify a few points regarding the method.
{}Firstly, one could object to the fact that $\delta$ is formally treated as 
small during the actual calculation and finally set to unity at the end. 
However, we recall that the only role attributed to this dummy parameter is 
to label the orders so that one can keep track of the extra diagrams which 
arise from the quadratic vertex $\delta \eta^2 \phi^2$.

{}Finally, one could ask how the LDE relates to other non-perturbative 
analytical methods such as the $1/N$ expansion. To see that, let us consider 
the same model discussed above for the case where the dynamical variables 
are a set of $N$ scalar fields, $\phi^a$  ($a=1,\ldots,N$). 
Proceeding as before 
the $\delta$ expansion would give the following result for 
${\cal E}$ 
at ${\cal O}(\delta)$,
\begin{equation}
{\cal E}^{(1)}(\eta)= \delta N \frac{1}{4}\eta + \delta \frac {N(N+2)}{3}
\frac{3\lambda}{16\eta^2} \;.
\label{nova}
\end{equation}
The application of the PMS to this quantity gives
\begin{equation}
{\cal E}^{(1)}({\bar \eta})=\left[\frac{(N+2)}{3}\lambda\right ]^{1/3}\; 
0.4290 \;,
\label{deln} 
\end{equation}
at 
\begin{equation}
{\bar {\eta}}=[1.5\;(N+2) \lambda ]^{1/3}\;.
\label{etan}
\end{equation}
Higher order contributions bring more factors of $N$ (more loops) making the 
calculation meaningless if $N$ is very large. However, this particular limit 
can also be properly handled by the LDE provided one defines $g=N\lambda$ 
declaring that the large-$N$ limit will be studied with fixed $g$ 
\cite{coleman}.
Using $g=N\lambda$ in Eq. (\ref {etan}) one sees that, in the large-$N$ limit, 
$\bar \eta $ is of order $N^0$, in terms of which Eq. (\ref {nova}) gives that 
${\cal E}^{(1)}({\bar \eta})$ is of order $N$, exactly as the leading $1/N$ 
result, as one can easily check.

An important result, proved in the context of the effective potential 
\cite {sunil}, shows that the LDE exactly reproduces large-$N$ results in 
any order in $\delta$ provided that one stays within the large-$N$ limit. 
Moreover, the LDE is sensitive to small-$N$ effects since these terms may 
appear in terms such as  $N(N+2)\lambda$ in Eq. (\ref {nova}). In fact,  
Ref. \cite {landau} shows how small-$N$ effects are effectively taken into 
account by the LDE in the context of the  $1+1$ dimensional Gross-Neveu model 
at finite temperatures where the results nicely converge, order by order, 
towards the exact result set by Landau's theorem.

The formal relationship in between the LDE and $1/N$ is investigated in 
detail in Refs. \cite {landau} and \cite {sunil}.
Here, we shall concern ourselves with the finite $N$ case only.

\section{The interpolated model for dilute homogeneous Bose gases}

Let us start by considering the typical model that describes a gas of
interacting boson particles, described by a complex scalar field $\psi$
with Lagrangian density given by  
\begin{eqnarray}
{\cal L}&= &\psi^*({\bf x},t)\left(  
i\frac{d}{dt}+\frac{1}{2m}\nabla^2\right)
\psi ({\bf x},t)
+\mu \psi^* ({\bf x},t) \psi ({\bf x},t) \nonumber \\
&-& \frac{1}{2}
\int ~ d^3 x' \psi({\bf x},t)\psi^*({\bf x},t)
V({\bf x}-{\bf x'})\psi({\bf x'},t)\psi^*({\bf x'},t) \; ,
\label{lagr}
\end{eqnarray}

\noindent
where $\mu$ is the chemical potential. Let us take the 
interatomic interaction potential as being the one 
for a hard sphere gas,

\begin{equation}
V({\bf x}-{\bf x'}) =  \frac{4 \pi a}{m} \delta ({\bf x}-{\bf x'})\;,
\end{equation}

\noindent
where $a$ is the $s$-wave scattering length.

We want to determine the deviation of the critical temperature $T_c$, of
the interacting model, in relation to the critical temperature for
Bose-Einstein condensation for a free gas, $T_{0}$, given by the usual
expression

\begin{equation}
T_0 =  \frac{2 \pi}{ m} \left(\frac{n}{\zeta(3/2)}  
\right)^\frac{2}{3}\;,
\label{T0}
\end{equation}

\noindent
where $n$ is the number density of the boson gas and $\zeta(3/2)\simeq 
2.612$.

As discussed in Refs. \cite{baymN} and \cite{arnold}, close to the
critical point we can reduce (\ref{lagr}) to an effective
three-dimensional model for the zero Matsubara frequency modes (the
static modes) of the fields $\psi$, given by the functional integration
of the non-zero modes, obtaining an effective action defined by
($\beta^{-1}=  T$)

\begin{equation}
\int_0^{\beta}d\tau \int d^3 x {\cal L}_{\rm Eucl}[\psi({\bf x},\tau), 
\psi^* ({\bf x},\tau)]
\rightarrow \beta \int d^3 x {\cal L}_{\rm eff} [\psi({\bf x}), 
\psi^* ({\bf x})]\;,
\label{Leff}
\end{equation}

\noindent
where ${\cal L}_{\rm Eucl}$ is the Lagrangian density in Euclidean space
($\tau = it$, as usual) and with the effective action for the static modes,
$\int d^3 x {\cal L}_{\rm eff}$,
being equivalent to a three-dimensional $O(2)$ field theory, 
defined by the action

\begin{equation}
S=  \int d^3x \left [ \frac {1}{2} | \nabla \phi |^2 + \frac {1}{2} r 
\phi^2 + \frac {u}{4!} (\phi^2)^2 \right ] \;\;,
\label{action O2}
\end{equation}

\noindent
where $\phi =  (\phi_1, \phi_2)$ is related to the original real
components of $\psi$ by $\psi_1=  (m  T)^{1/2} \phi_1$
and $\psi_2=  (m T)^{1/2} \phi_2$ while  $r$ and $u$
are given by

\begin{equation}
r=  - 2m \mu\;,\;\;\; 
u =  48 \pi a m T \;.
\end{equation}

\noindent 
By considering the usual interpolation prescription given by Eq. (\ref{int}) 
we write

\begin{equation}
S \rightarrow S_{\delta} =  \delta S + (1 - \delta) S_0 \;,
\end{equation}

\noindent
where $S_0$ is quadratically (exactly solvable) in the fields.

One can choose

\begin{equation}
S_0 =  \frac{1}{2}  \left [ | \nabla \phi|^2 + R \phi^2  \right ] \;,
\label{S0}
\end{equation}

\noindent 
where $R= r+\eta^2$, obtaining

\begin{equation}
S_{\delta}=  \int d^3x \left [ \frac {1}{2} | \nabla \phi |^2 + 
\frac {1}{2} R \phi^2  - \frac{\delta}{2} \eta^2 \phi^2 +  
\frac {\delta u}{4!} (\phi^2)^2 \right ] \;\;,
\label{Sdelta}
\end{equation}

\noindent
with $\eta$ being an arbitrary parameter, 
with mass dimensions, which is fixed
at a finite order in $\delta$ by the PMS condition, Eq. (\ref{pms1}). 
Here we will optimize
the physical quantity represented by $\langle \phi^2  
\rangle$ which, as we shall see, is directly related to the critical
temperature shift $\Delta T_c/T_c$. 
Let us first define the density for the interacting case

\begin{equation}
n =  mT \langle \phi^2 \rangle_u\;,
\label{density}
\end{equation}

\noindent
where, for the $O(N)$ symmetric model, $\langle \phi^2 \rangle_u$ is expressed 
in terms of the
three-dimensional dressed Green's function $G_\delta (p)$ as

\begin{equation}
\langle \phi^2 \rangle_u= \sum_{i= 1}^{N} \langle \phi^2_i \rangle_u =  N  
\int \frac {d^3 p}{(2 \pi)^3}
G_\delta (p)\;,
\label{phi2}
\end{equation}

\noindent
where

\begin{equation}
G_\delta (p)=  
\left[ p^2 + R - 
\delta{\eta^2} + \Sigma_{\delta}(p)  \right  ]^{-1}\;,
\end{equation}

\noindent
and $\Sigma_{\delta}(p)$ is the $\phi$ field renormalized 
self energy which will be evaluated perturbatively in powers of  
$\delta$. 

{}At the critical temperature the original system must 
exhibit infinite correlation length, which means that at $T_c$ and   
$\delta= 1$ (the original theory),
$G^{-1}_\delta (0) =  0$. Then, one gets the relation

\begin{equation}
r= -\Sigma_{\delta}(0)\;,
\label{HP}
\end{equation}

\noindent
which is just the form of the
Hugenholtz-Pines theorem. We must stress that the choice 
(\ref{Sdelta}) respects the Hugenholtz-Pines Theorem at all orders
in $\delta$.

Now, by using the relation (\ref{HP}) in (\ref{phi2}),
one can write 

\begin{equation}
\langle \phi^2 \rangle_u = 
\int \frac {d^3 p}{(2 \pi)^3}
\frac{N}{p^2 +\eta^2}\left [ 1  + \frac {(-\delta{\eta^2}) + 
\Sigma_{\delta}(p) - \Sigma_{\delta}(0) }{p^2 +\eta^2} \right ]^{-1}
\label{phi22}
\end{equation}

\section{Evaluation of $\langle \phi^2 \rangle$ to ${\cal O}(\delta^2)$}

Expanding the above expression, in powers of $\delta$,
to ${\cal O}(\delta)$ one sees that the only contribution to the self 
energy is a 
momentum independent tadpole diagram which is canceled  by the condition 
on $r$. 
Then, to order $\delta$, we obtain 

\begin{equation}
\langle \phi^2 \rangle_u =  \int \frac {d^3 p}{(2 \pi)^3} 
\frac{N}{p^2 +\eta^2}\left [
1+ \frac{\delta \eta^2}{p^2 +\eta^2}  \right ]\;,
\end{equation}

\noindent
which is $u$ independent and cannot furnish non-perturbative results. 
At next order in $\delta$ the only momentum dependent contribution to  
the self energy comes 
from the two loop  setting sun diagram, which is of order $\delta^2$. Then,  
we obtain

\begin{equation}
\langle \phi^2 \rangle_u =  \int \frac {d^3 p}{(2 \pi)^3}  
\frac{N}{p^2 +\eta^2}\left [1+ 
\frac{\delta\eta^2}{p^2 +\eta^2}+ \frac {\delta^2 \eta^4}{(p^2  
+\eta^2)^2} - 
\frac {\Sigma_{ss}(p) - \Sigma_{ss}(0)}{p^2 +\eta^2} 
\right ]\;,
\label{expansion}
\end{equation}

\noindent
where $\Sigma_{ss}(p)$ represents the setting sun contribution
to the self energy,

\begin{equation}
\Sigma_{ss}(p) =  - \frac{(N+2) u^2 \delta^2}{18} 
\int \frac {d^3 k}{(2 \pi)^3}  \frac {d^3 q}{(2 \pi)^3} 
\frac{1}{(k^2+\eta^2) (q^2 +\eta^2)
[(p+q+k)^2+\eta^2]} \;.
\end{equation}

Note that 
$\eta$ acts naturally as an infrared cutoff 
so we do  not have to worry about these type of divergences. 
The first three terms in Eq. (\ref{expansion}) represent one-loop
diagrams with different powers of 
$\delta\eta^2$ insertions. We regularize all diagrams with dimensional 
regularization in arbitrary dimensions 
$d= 3-2\epsilon$ and  carry the renormalization with the  
$\overline{MS}$ 
scheme. So the momentum integrals are replaced by

\[
\int \frac {d^3 p}{(2 \pi)^3} \to \int_p   
\equiv \left(\frac{e^{\gamma_E} M^2}{4 \pi} \right)^\epsilon
\int \frac {d^d p}{(2 \pi)^d} \;,
\]

\noindent
where $M$ is an arbitrary mass scale and $\gamma_E \simeq 0.5772$ is the 
Euler-Mascheroni constant. One then obtains the
${\cal O}(\delta^2)$ one loop contributions 

\begin{equation}
- \frac{N\eta}{4\pi}+\frac {\delta}{2} \frac{N\eta}{4\pi} + 
\frac{\delta^2}{8}\frac{N\eta}{4\pi} + {\cal O}(\epsilon)\;,
\label{first3}
\end{equation}

\noindent
where we have used the expression \cite{braaten}

\begin{equation}
\int_p \frac{1}{p^2 + \eta^2} = 
- \frac{\eta}{4 \pi} \left[1+ 2 \epsilon \left(\ln \frac{M}{2 \eta}
+1 \right) 
+ {\cal O}(\epsilon^2) \right]
\end{equation}

\noindent
and its derivatives with respect to $\eta^2$ to determine Eq. (\ref{first3}). 
The setting sun 
self energy diagram, with zero external momentum, is given by
(see, for example, Ref. \cite{braaten})

\begin{eqnarray}
\Sigma_{ss} (p)\Big|_{p= 0} =  - \frac{(N+2)}{18} 
\frac{u^2 \delta^2}{(8 \pi)^2} \left[
\frac{1}{\epsilon} + 4 \ln \frac{M}{2 \eta} +2 + 4 \ln\frac{2}{3} +
{\cal O}(\epsilon) \right] \;,
\end{eqnarray}

\noindent
from which one gets

\begin{equation}
\int_{p}\frac{N}{(p^2 +\eta^2)^2}\Sigma_{ss}(0)= 
- \frac{N(N+2)}{9}\frac{\delta^2 u^2}{ (8 \pi)^3 \eta}
\left [\frac{1}{2\epsilon} +3 \ln \left (\frac {M}{2\eta} \right ) + 1 +
2 \ln \frac{2}{3} + {\cal O}(\epsilon)\right]\;.
\label{ss0}
\end{equation}

The momentum dependent setting sun contribution can be written as

\begin{equation}
\int_{p}\frac{N}{(p^2 +\eta^2)^2}\Sigma_{ss}(p)= 
\frac{N (N+2) u^2 \delta^2}{72} \frac{d}{d \eta^2} I_{\rm bask}\;,
\label{ss}
\end{equation}

\noindent
where \cite {braaten}

\begin{eqnarray}
I_{\rm bask} &= & \int_{pkq} \frac{1}{(p^2+\eta^2) (k^2+\eta^2) (q^2  
+\eta^2)
[(p+q+k)^2+\eta^2]} \nonumber \\
&= &
- \frac{\eta}{(4 \pi)^3} \left[ \frac{1}{\epsilon} +
6 \ln \frac{M}{2 \eta} + 8 - 4 \ln 2 + {\cal O}(\epsilon) \right]\;.
\end{eqnarray}

\noindent
We then obtain for Eq. (\ref{ss})

\begin{equation}
\int_{p}\frac{N}{(p^2 + \eta^2)^2}\Sigma_{ss}(p)= 
-\frac{N(N+2)}{9}\frac{ \delta^2 u^2}{ (8 \pi)^3\eta}
\left [  \frac{1}{2 \epsilon}+3 \ln \left (\frac {M}{2\eta} \right ) +1
-2  \ln2 + {\cal O}(\epsilon) \right]\;.
\label{ssp}
\end{equation}

\section{The Temperature Shift in the Optimized Linear $\delta$ Expansion}

Using Eqs. (\ref{first3}), (\ref{ss0}) and (\ref{ssp}) in
(\ref{expansion}), we 
determine $\langle \phi^2 \rangle_u$ at order $\delta^2$.
Note that all divergences in $\epsilon$ cancel and 
that at order $\delta^2$, $\langle \phi^2 \rangle_u$ is
a finite quantity.
One can now set $\delta= 1$ and optimize $\langle \phi^2 \rangle_u$ 
with the PMS. After that one sets $u=0$ in the optimized  $\langle \phi^2 
\rangle_u$ obtaining the $\delta$ expansion result for the critical 
temperature shift \cite{baymprl,baymN,arnold}
 
\begin{equation}
\frac{\Delta T_c}{T_c} \simeq - \frac{2 m T_0}{3  n}
\Delta\langle \phi^2 \rangle=  - \frac{2 m T_0}{3  n}\left [\langle 
\phi^2\rangle_u- \langle \phi^2 \rangle_0 \right ] \;,
\label{deltaTc}
\end{equation}

\noindent
where $T_0$ is given by 
Eq. (\ref{T0}). 
At this stage it should be clear that it is preferable to optimize 
$\langle \phi^2 \rangle_u$ rather than
$\Delta \langle \phi^2 \rangle$ because the latter quantity is less  
$\eta$ dependent.
One then obtains

\begin{equation}
{\bar \eta} =  \pm \left [ \frac {6 (N+2) u^2 \ln \frac{4}{3} }{(36  
\pi)^2} 
\right ]^{1/2}\;,
\label{etabarra}
\end{equation}
which leads to
\begin{equation}
\langle \phi^2\rangle_u =  \mp \frac {uN}{192 \pi^2} 
\left [ 6(N+2)\ln\frac{4}{3} \right ]^{\frac{1}{2}}\;.
\label{result}
\end{equation}

\noindent
In principle one could not single out one solution in favor of the other  
and we must be careful in choosing the appropriate one.
Equation (\ref {result}) implies that the optimized $\langle \phi^2 \rangle_0$ 
vanishes no matter which sign is chosen. Now $\langle \phi^2 \rangle_0$  
represents the density (divided by a factor $m T$) in the absence of 
interactions, which turns 
out to be zero for the present effective theory. However, this is of no 
concern here since one is really interested in the difference 
$\langle \phi^2 \rangle_u - \langle \phi^2 \rangle_0$, 
represented by $\Delta T_c$. Also, one knows that the density of the 
interacting gas ($\langle \phi^2 \rangle_u$) should be smaller than that of 
the non interacting gas, 
which means that here one should have  $\langle \phi^2 \rangle_u < 0$,  
which is achieved by 
selecting the positive $\bar \eta$. Using this in 
Eq. (\ref{deltaTc}), we then get our final result,
 
\begin{equation}
\frac{\Delta T_c}{T_c} \simeq \frac{2 \pi}{\zeta(3/2)^{\frac{4}{3}}} 
\frac {N}{3} 
\left [ 6 (N+2)\ln \frac{4}{3} \right ]^{\frac{1}{2}} a  
n^{\frac{1}{3}}\;.
\label{deltaT}
\end{equation}

\noindent
Setting $N= 2$ in the above expression yields
\begin{equation}
\frac{\Delta T_c}{T_c} \simeq 3.059 \: a n^\frac{1}{3}\;.
\end{equation}

\noindent 
Using the $1/N$ expansion Baym, Blaizot and Zinn-Justin 
\cite{baymN} obtain $\Delta T_c/T_c \simeq 2.33 \: a n^{1/3}$ in the leading 
order while  Arnold and Tom\'asik \cite{arnold} obtain 
$\Delta T_c/T_c \sim 1.71  \:
an^{1/3}$ considering the next to leading order in the same approximation.
Our result is closer to $\Delta T_c/T_c \simeq 2.9 \; a n^{1/3}$ 
obtained in  Ref. \cite{baymprl} with a method which sums setting sun 
contributions in a self consistent way. These analytical results, including 
ours, are compared with the recent and earlier Monte Carlo estimates in 
Ref. \cite{arnold1}.
{}Finally let us remark that the result given by Eq. (\ref {deltaT}) is 
valid only for finite $N$. This can be understood as follows. In a large $N$
study one would have to consider $uN$ as fixed (meaning that $u \sim N^{-1}$) 
and so, by taking $N$ large in Eq. (\ref {expansion}), one sees that the 
setting sun diagrams of ${\cal O }(N^0)$  should be neglected since there 
are one 
loop 
diagrams of ${\cal O}(N)$. However, these terms are linear in $\eta$ as 
shown by 
their contribution, Eq. (\ref {first3}), and so the PMS does not give any 
meaningful result in this limit for the present model as opposed to the 
AO case.  The difference arises mainly because momentum independent tadpole 
diagrams of ${\cal O}(N)$ are now being subtracted due to 
the Hugenholz-Pines theorem while the one loop momentum independent diagrams 
survive. Then, the PMS generates nontrivial results 
only by mixing diagrams which would belong to different orders in a standard 
large-$N$ application. 
To have a rough idea about what is being summed one can consider the second 
and fourth terms in Eq. (\ref{expansion}) together with their integrated 
forms. Then it is clear that, apart from a numerical factor, the optimized 
${\bar \eta}^2$ given by Eq. (\ref {etabarra}) behaves as the optimized 
$\Sigma_{ss}(p) - \Sigma_{ss}(0)$. One can then see that, to this order,
the optimization dresses the simple propagator $(p^2+{\eta^2})^{-1}$, 
present in Eq. (\ref {expansion}), with setting sun features giving 
non-perturbative results which are compatible with the ones obtained in the  
self consistent summation of Ref. \cite {baymprl}.     

\section{Conclusions}
 
We conclude that our results for the critical temperature indeed
reproduce the expected behavior obtained from other studies, which is
$T_c \simeq T_0 (1 + \gamma a n^{1/3})$.
We obtain an analytical expression for the
numerical coefficient $\gamma$ in terms of finite values of $N$.
Our final numerical results are similar to the ones obtained with
 the self consistent summation \cite {baymprl} predicting that the
numerical value of $\gamma$ is greater than the ones predicted by the
$1/N$ expansion at leading order \cite {baymN} and next to leading order
\cite {arnold}. All these analytical results, including ours, have 
been compared with earlier and recent Monte Carlo results in Ref. 
\cite{arnold1}.
It should be clear that the present calculation has been carried out to 
an order where only one two loop diagram contributes and so the quality 
of the approximation is hard to be inferred from a quantitative point of view.
In fact, the purpose of the present application was just to introduce the 
method as a possible alternative to study the condensation problem. 
Nevertheless, one should remark that although carried out in a completely 
different fashion,  our simple application seems to capture much of the 
features of the self consistent calculation performed by Baym et al. 
\cite {baymprl}.

Also, our work does not exhaust
the different ways in which this method can be implemented within this particular 
problem and the possibility of further improvements is still wide open. 
This could be achieved by investigating alternative forms of implementing 
the method within this model, including an investigation of 
the best quantity to be optimized, and/or by pushing the calculation 
to higher orders. It is possible that with more refinements this method 
will generate even better numerical results with the advantage, as shown 
in the paper,  of being
considerable simpler and easier to use than
all previous methods used to determine the behavior of $T_c$.                 

Due to its
simplicity and easy implementation, we believe that the optimized
$\delta$ expansion can also be useful in other aspects of the
theoretical study and understanding of the Bose condensation of dilute
atomic gases, as determining the correct corrections to the energy
spectrum, or in applications related to the recent studies of the dynamics
of the Bose-Einstein condensate formation \cite{BFR}.
The results of
Ref. \cite{aodin} are particularly motivating in the context 
of applying the optimized linear $\delta$ expansion also to dynamical
problems. 
{}Finally, we point out that Bedingham and Evans \cite {ts} have successfully 
extended the present work 
to the ultra-relativistic case.

\acknowledgments

The authors thank Philippe Garcia for his careful reading of Sec. II and 
Franck Lalo\"e for 
interesting discussions regarding Bose-Einstein condensation
as well as the method employed in this work. F.F.S.C. and M.B.P. were 
partially supported by CNPq-Brazil.
R.O.R. was supported by CNPq-Brazil and SR2-UERJ.


\begin{references}

\bibitem{dalfovo} F. Dalfovo, S. Giorgini and L. P. Pitaevskii,
Rev. Mod. Phys. {\bf 71}, 463 (1999).

\bibitem{other} W. Ketterle, D. S. Durfee and D. M. Stamper-Kurn,
in {\it Bose-Einstein Condensation in Atomic Gases}, Proceedings of
the International School of Physics "Enrico Fermi", editors
M. I. Inguscio, S. Stringari and C. E. Wieman (IOS Press, Amsterdam
1999).

\bibitem{fetter} A. L. Fetter and J. D. Walecka, 
``Quantum Theory of Many-Particle Systems'', (McGraw Hill, 1971), section 28.

\bibitem{toyoda} T. Toyoda, Ann. Phys. (N.Y.) {\bf 141}, 154 (1982).

\bibitem{huang} K. Huang, Phys. Rev. Lett. {\bf 83}, 3770 (1999). 

\bibitem{gruter} P. Gr\"uter, D. Ceperley and F. Lalo\"e,
Phys. Rev. Lett. {\bf 79}, 3549 (1997).

\bibitem{holzmann-krauth} M. Holzmann and W. Krauth, Phys. Rev.
Lett. {\bf 83}, 2687 (1999).

\bibitem{markus} M. Holzmann, P. Gr\"uter and F. Lalo\"e, 
Euro. Phys J. {\bf B 10}, 739 (1999).

\bibitem{russos} N. Prokof'ev and B. Svistunov, cond-mat/0103149.

\bibitem{arnold1} P. Arnold and G. Moore, cond-mat/0103228.

\bibitem{baymprl} G. Baym, J.-P. Blaizot M. Holzmann, F. Lalo\"e 
and D. Vautherin, Phys. Rev. Lett. {\bf 83}, 1703 (1999).

\bibitem{baymN} G. Baym, J.-P. Blaizot and J. Zinn-Justin, Europhys. 
Lett. {\bf 49}, 150 (2000).

\bibitem{arnold} P. Arnold and B. Tom\'{a}sik, Phys. Rev. {\bf A 62}, 
063604 (2000).

\bibitem{stoof} M. Bijlsma and H. T. C. Stoof, Phys. Rev.
{\bf A 54}, 5085 (1996).


\bibitem{reppy} J. Reppy, B. Crooker, B. Hebral, A. Corwin,
J. He and G. Zassanhaus, Phys. Rev. Lett. {\bf 84}, 2060
(2000).


\bibitem{linear} A. Okopi\'nska, Phys. Rev. {\bf D 35},
1835 (1987); A. Duncan and M. Moshe, Phys. Lett. {\bf B 215},
352 (1988).


\bibitem{seznec}V. I. Yukalov, Mosc. Univ. Phys. Bull. {\bf 31}, 10 (1976).


\bibitem{seznec1}R. Seznec and J. Zinn-Justin, J. Math. Phys. {\bf 20},
1398 (1979).

\bibitem{seznec2}J. C. LeGuillou and J. Zinn-Justin, Ann. Phys. (N.Y.)
{\bf 147}, 57 (1983).

\bibitem{MR1}M. B. Pinto and R. O. Ramos, Phys. Rev. {\bf D 60}, 105005 
(1999); {\it ibid.} {\bf D 61}, 125016 (2000). 

\bibitem{hugh} H. F. Jones and P. Parkin, Nucl. Phys. {\bf B 594},
518 (2001).

\bibitem{okotem}A. Okopi\'{n}ska, Phys. Rev. {\bf D 36},
2415 (1987).

\bibitem{landau}S. K. Gandhi and M. B. Pinto, Phys. Rev {\bf D 49},
4258 (1994).

\bibitem{njlft}M. B. Pinto, Phys. Rev {\bf D 50},
7673 (1994).

\bibitem{ian} I. R. C. Buckley, A. Duncan and H. F. Jones, Phys. Rev. 
{\bf D 47}, 2554 (1993); A. Duncan and H. F. Jones, {\it ibid.} {\bf D 47},
2560  (1993); C. M. Bender, A. Duncan and H. F. Jones, {\it ibid.} {\bf D 49},
4219 (1994); C. Arvanitis, H. F. Jones and C. S. Parker,
Phys. Rev. {\bf D 52}, 3704 (1995). 

\bibitem{guida} R. Guida, K. Konishi and H. Suzuki, Ann. Phys. (N.Y.) 
{\bf 249}, 109 (1996).


\bibitem{gas} G. Krein, D. P. Menezes and M. B. Pinto, Phys. Lett. 
{\bf B 370}, 5 (1996);
G. Krein, R. S. M. Carvalho, D. P. Menezes, M. Nielsen, 
M. B. Pinto, Eur. Phys. J. {\bf A 1}, 45 (1998).


\bibitem{evans} T. S. Evans, H. F. Jones and A. Ritz, 
Nucl. Phys. {\bf B 517}, 599 (1998).

\bibitem{zj} R. Guida and J. Zinn-Justin, Nucl. Phys. {\bf B 489}, 626 (1997).

\bibitem{pms} P. M. Stevenson, Phys. Rev. {\bf D 23} 2916 (1981).

\bibitem{phil} B. Bellet, P. Garcia and A. Neveu, Int. J. of Mod. 
Phys {\bf A 11}, 5587 (1997); {\it ibid.} {\bf A 11}, 5607 (1997).


\bibitem{carl} C. M. Bender, K. Olausen and P. S. Wang, 
Phys. Rev. {\bf D 16} 1780 (1977).

\bibitem{bane} K. Banerjee, S.P. Bhatnager, V. Choudry and S.S. Kanwal, 
Proc. R. Soc. London {\bf A 360} 575 (1978).


\bibitem{aodin}H. F. Jones, P. Parkin and D. Winder, hep-th/0008069.

\bibitem{coleman} S. Coleman, ``{\it Aspects of Symmetry}'' 
(Cambridge University Press, Cambridge, 1988).

\bibitem{sunil}S. K. Gandhi, H. F. Jones and M. B. Pinto, 
Nucl. Phys. {\bf B359} 429 (1991).

\bibitem{braaten} E. Braaten and A. Nieto, Phys. Rev. {\bf D51}, 6990 
(1995).

\bibitem{BFR} D. G. Barci, E. S. Fraga and R. O. Ramos, Phys. Rev. Lett. 
{\bf 85}, 479 (2000).

\bibitem{ts} D. J. Bedingham and T. S. Evans, hep-ph/0011286.


\end{references}
\end{document}